# WIENER RECONSTRUCTION OF THE LARGE SCALE STRUCTURE

S. Zaroubi[1,2], Y. Hoffman[2], K.B. Fisher[3,4] and O. Lahav[4]

(1) Astronomy Department and Center for Particle Astrophysics, Campbell Hall, University of California, Berkeley, CA 97420, U.S.A.

(2) Racah Institute of Physics, The Hebrew University, Jerusalem 91904, Israel

(3) Institute for Advanced Study, Olden Lane, Natural Sciences, Bldg E, Princeton, NJ 08540, U.S.A.

(4) Institute of Astronomy, Madingley Rd., Cambridge CB3 OHA, UK

## ABSTRACT

The formalism of Wiener filtering is developed here for the purpose of reconstructing the large scale structure of the universe from noisy, sparse and incomplete data. The method is based on a linear minimum variance solution, given data and an assumed *prior* model which specifies the covariance matrix of the field to be reconstructed. While earlier applications of the Wiener filer have focused on estimation, namely suppressing the noise in the measured quantities, we extend the method here to perform both prediction and dynamical reconstruction. The Wiener filter is used to predict the values of unmeasured quantities, such as the density field in un-sampled regions of space, or to deconvolve blurred data. The method is developed, within the context of linear gravitational instability theory, to perform dynamical reconstruction of one field which is dynamically related to some other observed field. This is the case, for example, in the reconstruction of the real space galaxy distribution from its redshift distribution or the prediction of the radial velocity field from the observed density field.

When the field to be reconstructed is a Gaussian random field, such as the primordial perturbation field predicted by the canonical model of cosmology, the Wiener filter can be pushed to its fullest potential. In such a case the Wiener estimator coincides with the Bayesian estimator designed to maximize the *posterior* probability. The Wiener filter can be also derived by assuming a quadratic regularization function, in analogy with the 'Maximum Entropy' method. The mean field obtained by the minimal variance solution can be supplemented with constrained realizations of the Gaussian field to create random realizations of the residual from the mean.

*Subject Headings: cosmology: large scale structure of universe; analytical– methods: data analysis*

## 1. Introduction

Mapping the distribution of galaxies and their peculiar velocity field constitutes a major research area in modern astronomy setting both the observational and theoretical foundations of cosmology and, in particular, of large scale structure (LSS). The large scale galaxy distribution offers a probe of the early universe and the nature of the primordial perturbation field, and can be used to set strong constraints on the values of cosmological parameters (*cf.*, Dekel 1994 for a review and references therein). Astronomical observations give us however only incomplete and noisy information on the real universe, as is true in all experimental sciences. Any experimental measurement necessarily provides an incomplete description of the physical quantities under consideration, due to the finite accuracy of the measuring device. The study of LSS poses a unique astronomical problem, namely the obscuration by the gas and dust in the disk of the Galaxy, the so-called Zone of Avoidance (ZOA), which masks a non-negligible part of the sky. To summarize, the problem addressed here is one often encountered in physics in general, and in the study of LSS in particular, namely the reconstruction of an underlying field from noisy and incomplete observational data.

The canonical model of cosmology assumes that structure has grown out of small density perturbations via the process of gravitational instability. These perturbations are usually assumed to satisfy the statistics of Gaussian random fields (GRFs). The initial Gaussian field is preserved only in the linear regime; one expects deviations from a Gaussian field on scales where structure has evolved into the non-linear regime. Although, the reconstruction method presented in this paper does not rely on the assumption of Gaussianity, its fullest potential is achieved when working within the frame work of GRFs. Linear theory in LSS is not the only case in which Gaussian probability distribution function (PDF) is important. It can be shown that the small fluctuations from from a state of local thermodynamic equilibrium constitute a GRF (Landau and Lifshitz 1980). These thermodynamic considerations do not strictly apply to the case of gravitational instability in an expanding universe. Yet, on the technical level a GRF behaves as if it corresponds to a state of finite temperature thermal fluctuations.

In this paper, the problem of reconstruction is solved by assuming statistical *prior* information of the underlying field one is trying to measure. This falls into two different approaches, which at first glance seem to be quite unrelated. One is the well known Bayesian approach which quantifies the *posterior* probability of the model given the data in terms of the likelihood probability of the occurrence of the data given the model and the *a priori* probability of the correctness of the model (*i.e.*, the *prior*). A powerful application of Bayes' theorem is the Maximum Entropy (MaxEnt) algorithm, where the *prior* probability is formulated in terms of statistical entropy (Gull 1989; Skilling 1989). The other approach is that of linear estimation and prediction based on the principle of minimal variance from a given set of measurements (Wiener 1949). Here, estimation refers to the estimate of the field at a measured point while prediction refers to the estimate at an unmeasured point. The former is known in the literature as the Wiener filter (WF), however here we shall use the term in the broader sense of both estimation *and* prediction. The WF algorithm is well known and has been widely applied in many branches of physics and astronomy (*cf.*, the excellent review of Rybicki and Press, 1992). Currently, the WF is overlooked in favor of other methods, such as MaxEnt, particularly in the field of image processing. However, we show that for the particular case of reconstructing the LSS, the WF is indeed the 'optimal' tool best suited to handle such systems, and moreover it coincides with the Bayesian estimation in the case of GRFs.

A primary aim of any LSS reconstruction technique is to recover an estimate of the smooth underlying cosmological perturbation field from observed data. We put a special emphasis on the smoothness of the underlying field. Astronomical observations of LSS are done by measuring a certain property (*e.g.*, position, velocity) of a discrete object (galaxy, cluster), yet one is often interested in calculating a mathematically continuous (hereafter smooth) field. This can be considered as a mere mathematical convenience, but it has a physical motivation. First, the actual number of galaxies in a given volume is usually much larger than the number observed, in particular in magnitude limited samples where at large distances one just sees the decreasing tail of the luminosity function. The sparseness of the observed galaxy distribution usually necessitates smoothing with finite spatial resolution, i.e. by representing the galaxy distribution as a smooth field. Further still, many theories of LSS predict that the universe is dominated by dark matter made of particles whose individual masses are much smaller than a galactic mass. In such a case the mass density and velocity fields can be considered smooth. This underlying smooth field is sampled by the (often only bright) galaxies, which, in the lack of any detailed theory of galaxy formation, are usually considered to constitute a Poisson process superimposed on the matter density field (*cf.*, Scherrer and Bertschinger, 1991 for a formal treatment). This finite sampling introduces statistical uncertainties, commonly referred to as "shot" noise, into the analyses of redshift surveys. Shot noise is often further amplified by the need to correct for the selection criteria of a given sample, e.g., giving increased weight to more distant high luminosity galaxies in a flux-limited survey. Any reconstruction method aimed at recovering the true underlying density field must therefore yield robust and reliable results when the observations are corrupted by a high degree of statistical uncertainty.

We have recently started a collaborative effort of reconstructing the LSS using the WF method, as first reported by Hoffman (1994a) and Lahav (1994a). The method has been used recently to reconstruct the angular density field of galaxies in the 1.2 Jy *IRAS* survey (Lahav *et al.* 1994a, hereafter LFHSZ) and the angular maps of the CMB temperature fluctuations observed by COBE (Bunn *et al.* 1994; hereafter BFHLSZ). The angular reconstruction method was extended to three dimensions by Fisher *et al.* (1994, hereafter FLHLZ) and used to recover the real space density, velocity and gravitational potential fields from the redshift distribution of galaxies in the 1.2 Jy *IRAS* survey. The reconstruction presented in FLHLZ was based on a decomposition of the redshift space density field in spherical coordinates (spherical harmonics and spherical Bessel functions). A similar study utilizing a Cartesian space representation, is currently being conducted by Bistolas, Zaroubi and Hoffman (1994). A cosmographical analysis of these reconstructions has also been presented recently (Hoffman, 1994b; Lahav, 1994b; LFHSZ). Given the substantial use and application of the WF reconstruction we present here a unified presentation of the method.

Many of the ideas presented in the present work are not fundamentally new, in particular the basic WF theory. However, until recently it has not been applied to the reconstruction of the LSS, and in particular, within the framework of GRFs. Our primary aim is to present a comprehensive and self-contained description of the WF algorithm. An overview of the linear inversion problem and its solution by the WF is given in §2. The WF is developed for various functional representations in §3. The use of WF for dynamical mapping purposes is presented in §4. Possible extensions of WF are discussed in §5. The paper concludes with a general discussion (§6).

## 2. Theory

### 2a. The Inversion Problem

Consider the case of a set of observations, or measurements, performed on an underlying field $\mathbf{s} = \{s_\alpha\}$ ($\alpha = 1, \ldots, N$), or on any field linearly related to $\mathbf{s}$, which yields a set of results or data points, $\mathbf{d} = \{d_i\}$ ($i = 1, \ldots, M$). Here we are interested in measurements that can be modeled mathematically as a linear convolution or mapping of the underlying field,

$$\mathbf{d} = \mathbf{R}\,\mathbf{s} + \boldsymbol{\epsilon}, \qquad (2-1)$$

where $\mathbf{R}$ is an $M \times N$ matrix which represents a response or point spread function (hereafter RF) and $\boldsymbol{\epsilon} = \{\epsilon_i\}$ ($i = 1, \ldots, M$), is the statistical uncertainty vector associated with the data. For simplicity, and with no loss of generality, we assume a discrete modeling of the field in some representation (e.g., a finite number of Fourier coefficients, spherical harmonics, or real space positions). The RF can represent the response of the measuring device (or procedure) to the underlying field. Commonly it represents the blurring, or smoothing, introduced by the measurement. The notion of a RF can however be extended to represent a theoretical relationship between two fields (e.g., a matrix which relates the Fourier coefficients of the density field to those of the velocity field (FLHLZ)).

Usually, the statistical uncertainties in the data arise due to instrumental response (*e.g.*, the noise in the individual COBE DMR pixels (BFHLSZ)). However, there are some cases in which the statistical uncertainties arise not from the measurement process but are intrinsic to the the underlying field. An example of such a quantity, which is of special significance in cosmology, is the shot noise contribution to the the galaxy distribution which arises from the sampling of the continuous density field by a finite number of galaxies. In such a case, it is more illuminating to write the noise contribution as $\boldsymbol{\epsilon} = \mathbf{R}\boldsymbol{\sigma}$ where $\boldsymbol{\sigma}$ is the shot noise the underlying field which would be measured if $\mathbf{R} = I$, the unity matrix. Smoothing or weighting procedures can introduce additional correlations in the statistical noise $\boldsymbol{\epsilon}$ (Scherrer and Bertschinger 1991). In any case we would like to emphasize that the correlation of the statistical uncertainties differs vastly from one case to another and it should be determined according to the problem at hand.

Mathematically, the act of reconstruction amounts to solving Equation 2-1 for **s** given **d**, where **R** is assumed to be known and the uncertainties are only statistically known. A naive approach to the problem is a deconvolution of the operator **R** by direct inversion, i.e., $\mathbf{s} = \mathbf{R}^{-1}\mathbf{d}$. This approach is, at best, far from optimal for several reasons. First, one often encounters situations where the number of independent data points is much smaller than the number of degrees of freedom of the field ($N \gg M$). In such case the data do not contain enough information to constrain all the underlying field degrees of freedom. Second, it is well known that the direct inversion will greatly amplify the statistical noise $\epsilon$ and can lead to an unstable deconvolution even when $M \sim N$. Due to these potential pitfalls of direct inversion, one is forced to resort to a regularization technique for solving Equation 2-1, *i.e.*, a method which will recover a reliable estimate of **s** given the data set **d**; as we see in the next section, this is precisely what the Wiener filter is designed to do.

## 2b. Wiener Filter

The canonical model of LSS assumes that the primordial perturbations constitute a random field, hence we adopt this assumption for the reconstruction technique introduced here. Our *prior* model assumes knowledge of the first two moments of the field, **s**, we wish to recover: namely its mean, $\langle \mathbf{s} \rangle$ (taken in what follows to be 0 for simplicity), and its covariance matrix,

$$\mathbf{S} = \left\langle \mathbf{s}\,\mathbf{s}^\dagger \right\rangle \equiv \left\{ \left\langle s_i\, s_j^* \right\rangle \right\} \quad . \tag{2-2}$$

In equation 2-2 and in what follows $\mathbf{s}^\dagger$ denotes the complex conjugate of the transpose of the vector **s** and $\langle ... \rangle$ denotes an ensemble average. Notice that no assumption has been made regarding the actual functional form of the probability distribution function (PDF) which governs the random nature of the field besides its first two moments. We define an optimal estimator of the underlying field, $\mathbf{s}^{\mathrm{MV}}$ (hereafter MV estimator), as the linear combination of the data, **d**, which minimizes the variance of the discrepancy between the estimator and all possible realizations of the underlying field. Thus one writes

$$\mathbf{s}^{\mathrm{MV}} = \mathbf{F}\,\mathbf{d}, \qquad (2-3)$$

where the $\mathbf{F}$ is an $N \times M$ matrix chosen to minimize the variance of the residual $\mathbf{r}$ defined by

$$\left\langle \mathbf{r}\,\mathbf{r}^\dagger \right\rangle = \left\langle (\mathbf{s} - \mathbf{s}^{\mathrm{MV}})(\mathbf{s}^\dagger - \mathbf{s}^{\mathrm{MV}\,\dagger}) \right\rangle. \qquad (2-4)$$

Carrying out the minimization of equation 2-4 with respect to $\mathbf{F}$ one finds the so-called WF,

$$\mathbf{F} = \left\langle \mathbf{s}\,\mathbf{d}^\dagger \right\rangle \left\langle \mathbf{d}\,\mathbf{d}^\dagger \right\rangle^{-1}. \qquad (2-5)$$

The MV estimator of the underlying field is thus given by

$$\mathbf{s}^{\mathrm{MV}} = \left\langle \mathbf{s}\,\mathbf{d}^\dagger \right\rangle \left\langle \mathbf{d}\,\mathbf{d}^\dagger \right\rangle^{-1} \mathbf{d}. \qquad (2-6)$$

The variance of the residual of the $\alpha$-th degree of freedom can be shown to be

$$\left\langle |r_\alpha|^2 \right\rangle = \left\langle |s_\alpha|^2 \right\rangle - \left\langle s_\alpha \mathbf{d}^\dagger \right\rangle \left\langle \mathbf{d}\,\mathbf{d}^\dagger \right\rangle^{-1} \left\langle \mathbf{d}\, s_\alpha^* \right\rangle. \qquad (2-7)$$

The noise term $\boldsymbol{\epsilon}$ is assumed to be statistically independent of the underlying field ($\left\langle \boldsymbol{\epsilon}\,\mathbf{s}^\dagger \right\rangle = 0$) and therefore the correlation matrices appearing in equation 2-6 follow directly from equation 2-1:

$$\left\langle \mathbf{s}\,\mathbf{d}^\dagger \right\rangle = \left\langle \mathbf{s}\,\mathbf{s}^\dagger \right\rangle \mathbf{R}^\dagger \equiv \mathbf{S}\,\mathbf{R}^\dagger \qquad (2-8)$$

and

$$\left\langle \mathbf{d}\,\mathbf{d}^\dagger \right\rangle \equiv \mathbf{D} = \mathbf{R}\,\mathbf{S}\,\mathbf{R}^\dagger + \left\langle \boldsymbol{\epsilon}\,\boldsymbol{\epsilon}^\dagger \right\rangle. \qquad (2-9)$$

For the case in which $\boldsymbol{\epsilon}$ is expressed in terms of $\boldsymbol{\sigma}$ one gets,

$$\mathbf{N}_\epsilon \equiv \left\langle \boldsymbol{\epsilon}\,\boldsymbol{\epsilon}^\dagger \right\rangle = \mathbf{R}\left\langle \boldsymbol{\sigma}\,\boldsymbol{\sigma}^\dagger \right\rangle \mathbf{R}^\dagger \equiv \mathbf{R}\,\mathbf{N}_\sigma\,\mathbf{R}^\dagger, \qquad (2-10)$$

$\mathbf{N}_\epsilon$ and $\mathbf{N}_\sigma$ are the correlation matrices of the noise $\epsilon$ and $\sigma$ respectively ($\mathbf{N}_\epsilon$ and $\mathbf{N}_\sigma$ are not necessarily diagonal). With these definitions, the expression for WF given in equation $(2-5)$ becomes

$$\mathbf{F} = \mathbf{S}\mathbf{R}^\dagger(\mathbf{R}\,\mathbf{S}\,\mathbf{R}^\dagger + \mathbf{N}_\epsilon)^{-1} \qquad (2-11a)$$

or

$$\mathbf{F} = \mathbf{S}(\mathbf{S} + \mathbf{N}_\sigma)^{-1}\mathbf{R}^{-1} \qquad (2-11b)$$

Although, equations 2-11a and 2-11b are mathematically equivalent, equation 2-11a is often more practical computationally since it requires only a single matrix inversion.[†], However, if $\mathbf{S}$ and $\mathbf{N}_\sigma$ are both diagonal, then equation 2-11b becomes easier to deal with numerically (*e.g.*, LFHSZ). Furthermore, equation 2-11b shows explicitly the two fundamental operations of the WF:[*] *inversion* of the RF operating on the data ($\mathbf{R}^{-1}$) and *suppression* the shot noise roughly by the ratio of $\frac{prior}{prior+noise}$ (if $\mathbf{S}$ and $\mathbf{N}$ are diagonal). Note that this ratio is less than unity, and therefore the method can not be used iteratively as successive applications of the WF would drive the recovered field to zero.

The variance of the residual given in equation 2-4 can be calculated easily using equation 2-11b. This calculation gives,

$$\left\langle \mathbf{r}\,\mathbf{r}^\dagger \right\rangle = \mathbf{S}\,(\mathbf{S}\,+\,\mathbf{N}_\sigma)^{-1}\mathbf{N}_\sigma.$$

In the rest of the paper we will consider the case where the uncertainties are expressed explicitly in the observational domain and the uncertainty matrix is assumed to be $\mathbf{N} = \mathbf{N}_\epsilon$.

## 2c. Conditional Probability

---

[†] In general, the matrices are not square; in these cases inversion refers to the pseudo-inverse, e.g. as defined in terms of Singular Value Decomposition (cf. § 5a.1).

[*] Some authors refer to the ratio, (prior/prior+noise), as the WF. However, it is not always possible to separate it from $\mathbf{R}^{-1}$; consequently our notation WF contains both the operations noise suppression and inversion of the response function.

We now consider the case where the *prior* model is extended to have a full knowledge of the random nature of the underlying **s** field, which is mathematically represented by the PDF of the field, $P(\mathbf{s})$. Knowledge of the measurement, sampling and selections effects implies that the joint PDF, $P(\mathbf{s},\mathbf{d})$, can be explicitly written. The conditional mean value of the field given the data can serve as an estimator of **s**,

$$\mathbf{s}^{\mathrm{mean}} = \int \mathbf{s} P(\mathbf{s}|\mathbf{d})\,\mathrm{d}\mathbf{s}. \qquad (2-12)$$

The standard model of cosmology assumes that the primordial perturbation field is Gaussian, and therefore on large scales where the fluctuations are still small the present epoch perturbations field will be very close to Gaussian. The statistical properties of the GRF depend only on its two-point covariance matrix; in particular the PDF of the underlying field is a multivariate Gaussian distribution,

$$P(\mathbf{s}) = \frac{1}{\left[(2\pi)^N \det(\mathbf{S})\right]^{1/2}} \exp\left(-\frac{1}{2}\mathbf{s}^\dagger \mathbf{S}^{-1}\mathbf{s}\right), \qquad (2-13)$$

determined by the covariance matrix **S** (Bardeen *et al.* 1986).

Now, if the noise is an independent GRF, then the joint PDF for the signal and data is,

$$P(\mathbf{s},\mathbf{d}) = P(\mathbf{s},\boldsymbol{\epsilon}) = P(\mathbf{s})P(\boldsymbol{\epsilon}) \propto \exp -\frac{1}{2}\left(\mathbf{s}^\dagger \mathbf{S}^{-1}\mathbf{s} + \boldsymbol{\epsilon}^\dagger \mathbf{N}^{-1}\boldsymbol{\epsilon}\right) \qquad (2-14),$$

while the conditional PDF for the signal given the data is the shifted Gaussian,

$$P(\mathbf{s}|\mathbf{d}) = \frac{P(\mathbf{s},\mathbf{d})}{P(\mathbf{d})} \propto P(\mathbf{s})P(\boldsymbol{\epsilon}) \propto \exp\left[-\frac{1}{2}\left(\mathbf{s}^\dagger \mathbf{S}^{-1}\mathbf{s} + (\mathbf{d}-\mathbf{R}\mathbf{s})^\dagger \mathbf{N}^{-1}(\mathbf{d}-\mathbf{R}\mathbf{s})\right)\right]. \qquad (2-15)$$

Note also that the second term in the exponent, in equation 2-15, is $-\frac{1}{2}$ the classical $\chi^2$ distribution. Following RP and Bertschinger (1987) we rewrite equation (2-15) by completing the square for **s**:

$$P(\mathbf{s}|\mathbf{d}) \propto$$
$$\exp\left[-\frac{1}{2}\left(\mathbf{s} - \mathbf{S}\mathbf{R}^\dagger(\mathbf{R}\mathbf{S}\mathbf{R}^\dagger + \mathbf{N})^{-1}\mathbf{d}\right)^\dagger\left(\mathbf{S}^{-1} + \mathbf{R}^\dagger \mathbf{N}^{-1}\mathbf{R}\right)\left(\mathbf{s} - \mathbf{S}\mathbf{R}^\dagger(\mathbf{R}\mathbf{S}\mathbf{R}^\dagger + \mathbf{N})^{-1}\mathbf{d}\right)\right].$$
$$(2-16)$$

The integral of equation 12 is trivially calculated now to yield $\mathbf{s}^{\mathrm{mean}} = \mathbf{s}^{\mathrm{MV}}$, the residual from the mean coincides with $\mathbf{r}$, which is Gaussian distributed with a zero mean and whose covariance matrix is $\left(\mathbf{S}^{-1} + \mathbf{R}^{\dagger}\mathbf{N}^{-1}\mathbf{R}\right)^{-1}$. The important result is that for GRFs the WF minimal variance reconstruction coincides with the conditional mean field. Maximizing equation 2-15 with respect to the field $\mathbf{s}$, yields yet another estimator, namely the maximum *a posteriori* estimate (MAP) of the field; it is easily shown that the MAP estimator coincides with the WF and conditional mean field i.e., $\mathbf{s}^{\mathrm{MV}} = \mathbf{s}^{\mathrm{mean}} = \mathbf{s}^{MAP}$.

In several cases one might be interested in creating a random realizations of a GRF which reproduces a given set of data points, the so-called technique of constrained realizations (CR) (Hoffman and Ribak 1991, hereafter HR). The idea of the CR method is to supplement the mean field (which is uniquely determined by the data and the *prior*) with a random realization of the residuals from the mean (see Bertschinger 1987, Rybicki and Press 1992 for a general discussion of CR). The resulting CR can be used as initial conditions for a non-linear N-body simulation or in a Monte Carlo study of the reconstructed field.

Within the framework of GRFs and a given *prior*, one can show that any realization can be split into two parts: the mean field which is determined by the constraints (data) and the residual field which is a GRF whose variance is given in Eq. 2-7. The main point of the CR is that the statistics of the residual field is independent of the actual numerical values of the constraints imposed (HR). Thus, one can make a random realization of the underlying field, based on the assumed *prior* model, and sample it in the same way the actual data is obtained (cf., equation 2-1),

$$\tilde{\mathbf{d}} = \mathbf{R}\tilde{\mathbf{s}} + \tilde{\boldsymbol{\epsilon}} \qquad (2-17)$$

Here, $\tilde{\mathbf{s}}$ and $\tilde{\boldsymbol{\epsilon}}$ are random realizations of the underlying field and the statistical uncertainties, respectively. A CR of the field given the data is given by (HR):

$$\mathbf{s} = \tilde{\mathbf{s}} + \mathbf{F}(\mathbf{d} - \tilde{\mathbf{d}}). \qquad (2-18)$$

Note that the variance of the residual obtained here is consistent with the expression derived above (equation 2-7).

Another estimator can be formulated from the point of view of Bayesian statistics. The main objective of this approach is to calculate the *posterior* probability of the model given the data, which is written according to Bayes' theorem as $P(\text{model}|\text{data}) \propto P(\text{data}|\text{model})P(\text{model})$. The estimator of the underlying field (*i.e.*, model, in Bayes' language) is taken to be the one that maximizes $P(\text{model}|\text{data})$, which is the most probable field. The Bayesian *posterior* PDF is given by:

$$P(\mathbf{s}|\mathbf{d}) \propto P(\mathbf{s})P(\mathbf{d}|\mathbf{s}), \qquad (2-19)$$

now, in the general case which is given by equation 2-1, where the *prior* assumed to be a Gaussian, equation (2-19) give:

$$P(\mathbf{s}|\mathbf{d}) \propto \exp -\frac{1}{2}\left(\mathbf{s}^\dagger \mathbf{S}^{-1}\mathbf{s} + (\mathbf{d}-\mathbf{R}\mathbf{s})^\dagger \mathbf{N}^{-1}(\mathbf{d}-\mathbf{R}\mathbf{s})\right) \qquad (2-20)$$

the Bayesian estimator, as it corresponds to the most probable configuration of the underlying field given the data and Gaussian *prior*, coincides with the $\mathbf{s}^{\text{MV}}$.

### 3. Representations

#### 3a. 'Orthogonality' and functional Basis' Sets

Cosmic dynamical fields (e.g., density, velocity and gravitational potential) can, like any mathematically well behaved fields, be expanded in any complete set of basis functions. Such a basis defines a space in which all of the physical, statistical and geometrical properties of the field are retained. Spaces of special interest for WF applications are those in which both $\langle \mathbf{s}\,\mathbf{s}^\dagger \rangle$ and $\langle \boldsymbol{\epsilon}\,\boldsymbol{\epsilon}^\dagger \rangle$ are diagonal (diagonal $\langle \mathbf{d}\,\mathbf{d}^\dagger \rangle$); in such a basis the WF has a particular simple form. Such a basis, if it exists, is determined by the homogeneity and isotropy of the underlying field and shot noise. We will refer to the property of bases which diagonalize the correlation matrices as 'statistical orthogonality'.

As a simple example, consider the Fourier decomposition of the density field derived from a full sky discrete galaxy catalogue. In the framework of linear gravitational instability the Fourier functional basis is commonly used for the following reasons: a) The density field is statistically 'orthogonal', namely, $\langle \delta_{\mathbf{k}}\, \delta_{\mathbf{k}'} \rangle = P(k)\delta_D(\mathbf{k}-\mathbf{k}')$. b) The function $e^{i\mathbf{k}\mathbf{r}}$ used in Fourier transform is an eigenfunction both of the $\nabla$ and the Laplacian operators, therefore, Fourier space maintains the correspondence between dynamical fields algebraically. In this simple case, the underlying and observed field are related by

$$\Delta_{\mathbf{k}} = \delta_{\mathbf{k}} + \epsilon, \qquad (3-1)$$

where $\delta_{\mathbf{k}}$ and $\Delta_{\mathbf{k}}$ represent the Fourier coefficients of the underlying and observed fields respectively, and $\epsilon$ (which is independent of $k$) is the transform of the shot noise term.

Eq. 3-1 is a special form of the general inverse equation 2-1 with $\mathbf{R} = \mathbf{I}$. The WF reconstruction is particularly simple and is given by (cf., equation 2-11b)

$$\delta_{\mathbf{k}}^{\mathrm{MV}} = \left( \frac{P(k)}{P(k) + \langle |\epsilon|^2 \rangle} \right) \Delta_{\mathbf{k}} \quad . \qquad (3-2)$$

Note that the WF is given by the ratio of power to (power + noise) and that in the limit of very good data $(P(k) \gg \langle |\epsilon|^2 \rangle)$ the WF tends to unity and one recovers the raw data. While, in the limit of bad data, $(P(k) \ll \langle |\epsilon_{\mathbf{k}}|^2 \rangle)$, the WF attenuates the signal in an effort to suppress the statistical noise. This simple example demonstrates a generic feature of the WF; when the data is of high signal to noise it is equivalent to direct inversion and when the data is poor it yields the null estimator.

### 3b. SH and Mask Inversion - Two Dimensional

Another convenient functional basis is the spherical harmonics and spherical Bessel functions basis, which is suitable for expanding the field at hand around a given point. It is easy to show that in such representation homogeneous and isotropic random fields are 'statistically orthogonal'.

Spherical harmonics (SH) have been used to probe the large scale structure from wide angle galaxy surveys (Peebles 1973, 1980; Scharf *et al.* 1992; Scharf & Lahav 1993). These analyses consist of expanding the angular galaxy distribution in a set of spherical harmonics which form an orthonormal functional basis when the expansion is carried out over the full, $4\pi$, sky. Galaxy surveys suffer from two basic problems, namely a significant shot noise and incomplete sky coverage. Now, in the absence of noise the full sky harmonics can be easily found by a matrix inversion (Scharf *et al.* 1992) but the shot noise destabilizes the inversion. The reconstruction of the *IRAS* angular structure is well suited for a WF approach, as indeed has been recently worked out by LFHSZ. Here we expand the brief description of LFHSZ and study the potential and limitations of the application of the WF to the angular structure. The method is applied here to mock *IRAS* catalogs, compiled from N-body simulations, to test its ability to reconstruct the underlying large scale structure.

Consider an underlying angular density field, given in terms of its spherical harmonic expansion,

$$a_{lm} = \int d\hat{\mathbf{r}}\, \mathcal{S}(\hat{\mathbf{r}})\, Y^*_{lm}(\hat{\mathbf{r}}). \qquad (3-3)$$

where the projected surface density is given by

$$\mathcal{S}(\hat{\mathbf{r}}) = \sum_{lm} a_{lm} Y_{lm}(\hat{\mathbf{r}}) \quad . \qquad (3-4)$$

This field is sampled by a finite distribution of galaxies, which suffers basically from incomplete sky coverage. This sampling yields a raw estimator of $a_{lm}$ by:

$$c_{lm} = \sum_{i}^{N_{gal}} Y^*_{lm}(\hat{\mathbf{r}}_i) \qquad (3-5)$$

The observed $c_{lm}$ and the underlying $a_{lm}$ are related by

$$c_{lm} = \sum_{l'm'} W^{mm'}_{ll'} a_{l'm'} + \epsilon_{lm} \qquad (3-6)$$

where
$$W_{ll'}^{mm'} = \sum_{i}^{N_{gal}} Y_{lm}^*(\hat{\mathbf{r}}_i) Y_{l'm'}^*(\hat{\mathbf{r}}_i), \qquad (3-7)$$
is the harmonics expansion of the mask and $\epsilon_{lm}$ is the shot noise term. Note that the mask introduces mode-mode coupling between the 'observed' harmonics even in the case where the underlying harmonics $a_{lm}$ are statistical independent. In the case of full sky coverage the mask converges to $W_{ll'}^{mm'} = \delta_l^{l'} \delta_m^{m'}$. The extraction of the monopole term, corresponds to the mean surface number density, should take into account that it is coupled to the other multipoles. (See Peebles 1973, 1980 for rigorous treatment of a non-zero monopole term.) The representation in terms of spherical harmonics assumes that the sum in equation 3-4 extends to an infinite harmonic number, $l$. In practice, one should check for convergence of the summation as $l$ goes to infinity. A cutoff at a maximal $l_{max}$ implies a filter which cuts the small scale power on an effective angular scale of $\sim \pi/l_{max}$.

Eq. 3-6 can be rewritten as
$$c_{lm} = \sum_{l'm'} W_{ll'}^{mm'} \{a_{l'm'} + \sigma_{l'm'}\} \qquad (3-8)$$
where $\sigma_{l'm'}$ is the shot noise term in the case of full sky coverage. Assuming that the shot noise behind the masked region is the same as in the other regions, it follows that $\langle \sigma_{lm} \, \sigma_{l'm'} \rangle = \mathcal{N} \delta_l^{l'} \delta_m^{m'}$ (Scharf et al. 1992), where $\mathcal{N}$ is the two dimensional projected average number density.

The recovery of the 'true' $a_{lm}$ from equation (3-8) amounts to a deconvolution or inversion, which in the presence of noise is well known to be unstable. The WF is invoked here to stabilize this deconvolution, thus serving two purposes, namely estimation and prediction. The WF yields an estimator of the underlying field suppressed off the shot noise, and a prediction of the galaxy distribution in the masked regions.

Substituting $\mathbf{c}$ from equation 3-8 in equation 2-6 one obtains the MV estimator of the underlying field,
$$\mathbf{a}^{\mathrm{MV}} = \langle \mathbf{a}\,\mathbf{a}^\dagger \rangle \, \langle (\mathbf{a}+\boldsymbol{\sigma})(\mathbf{a}^\dagger + \boldsymbol{\sigma}^\dagger) \rangle^{-1} \mathbf{W}^{-1} \mathbf{c}$$

$$=diag\left\{\frac{A_l}{A_l+\mathcal{N}}\right\}\mathbf{W}^{-1}\mathbf{c}\equiv\mathbf{B}^{-1}\mathbf{C}, \qquad (3-9)$$

where for a statistical isotropic perturbation field the covariance matrix of the $a_{lm}$ is diagonal, $\langle a_{lm}a_{lm}^*\rangle = A_l\delta_{ll'}\delta_{mm'}$. Here the data and estimator vectors, $\mathbf{C}$ and $\mathbf{a}^{\mathrm{MV}}$, are assumed to be of the same length and the $\mathbf{B}^{-1}$ is a square matrix. In the case of a full sky coverage, where $\mathbf{W}$ is the unity matrix the WF corresponds to a diagonal isotropic (no $m$ dependence) matrix which affects only the amplitude but not the 'phases' of the harmonics. In the case where the observation is dominated by noise the WF approaches the null matrix and the estimator vanishes. In the limit of negligible noise but partial sky coverage the WF corresponds to a straightforward deconvolution of the 'observed' $c_{lm}$. The variance of the residual from the mean is given by:

$$\langle[\mathbf{a}-\mathbf{a}^{\mathrm{MV}}][\mathbf{a}^\dagger-\mathbf{a}^{\mathrm{MV}^\dagger}]\rangle = diag\left\{\frac{\langle\sigma_a^2\rangle A_l}{A_l+\langle\sigma_a^2\rangle}\right\}, \qquad (3-10)$$

where it is dominated by the theoretical power spectrum when $\langle\sigma_a^2\rangle \gg \langle a_l^2\rangle_{th}$, and vanishes if the signal to noise ratio is very big.

Here we apply this method to a standard CDM simulation characterized by $\Omega_0 h = 0.5$ with particles selected to mimic the $IRAS$ 1.2Jy galaxy catalogue. The simulation evolves the particles until the rms variance of the density field in a sphere of $8Mpc/h$ reached $\sigma_8 = 0.62$ (see Fisher, Scharf and Lahav 1994, for more details). At the first stage we assume a full sky coverage for which the coupling matrix, $\mathbf{W}$, is the unity matrix. Figure 1 shows the harmonic reconstruction of the projected counts of the simulation, using the raw all-sky coefficients $c_{lm}$ up to $l_{max} = 15$ of the Mock $IRAS$ catalogue in Galactic Aitoff projection. Although, in this case the sky is fully covered, WF is used to suppress the shot noise giving the most probable underlying density field. Note, that in this case WF changes only the harmonics amplitudes and do not affect the relative phases. Figure 2 shows the Wiener reconstruction of the all-sky map of Figure 1, using the Wiener filtered coefficients, $a_{lm}$, up to $l_{max} = 15$, where a standard CDM power spectrum is assumed and the shot-noise is taken from the simulation (392 galaxies per steradian). Figure 3 shows the Wiener factor (the ratio $\frac{A_l}{A_l+\langle\sigma_a^2\rangle}$) versus the harmonic number $l$, this factor drops monotonically with $l$ due to the increasing of the relative shot noise power ($\sigma_a^2/A_l$) with $l$. An application of the method for projected Mock $IRAS$ 1.2 Jy catalogue with a ZOA $|b| = 15°$ is presented in §5a.1 and for the real $IRAS$ data with $|b| = 5°$ at LFHSZ.

A similar approach has been adopted recently by BFHLSZ to analyze the $\Delta T/T$ map detected by COBE's DMR. Another basis that can be used to produce WF estimator of the CMB distribution, is the one introduced by Górski (1994). In this basis the inner product is defined by an integral over the incomplete part of the sky covered by the radiation. Orthogonality is obtained by applying Gramm-Schmidt orthogonalization method on the basis of spherical harmonics.

### 3c. SH and Spherical Bessel Functions for 3D Reconstruction

Here we use spherical harmonics and spherical Bessel functions, $j_l(kr)$ (cf., Arfken 1985; Jackson 1975), to expand the three dimensional galaxy distribution in whole-sky surveys. Recently, FLHLZ (see also Heavens and Taylor 1994) used this basis to probe the underlying density, velocity and potential fields, from $IRAS$ 1.2 Jy redshift catalogue, where they used WF for both noise suppression and deconvolution of the redshift distortion. FLHLZ neglected the coupling introduced by the incomplete sky coverage, which is a fair assumption for $IRAS$ catalogue. Here we use the same field to expand galaxy survey which suffers both from incomplete sky coverage and selection effect, where the redshift distortion is neglected.

Here we assume an infinite sample, hence, $k$ will be continuous (for detail discussion on the effect of finite sample and its boundary conditions see FLHLZ). The expansion is defined as follows:

$$\bar{\rho}\,\delta(\mathbf{r}) = \frac{2}{\pi} \sum_{lm} Y_{lm}(\hat{\mathbf{r}}) \int_0^\infty \rho_{lm}(k) j_l(kr) k^2 dk \qquad (3-11)$$

where the expansion coefficients $\rho_{lm}(k)$ are defined as

$$\rho_{lm}(k) = \int \delta(\mathbf{r}) j_l(kr) Y_{lm}(\hat{\mathbf{r}}) d^3 r \qquad (3-12)$$

The relation between the Fourier components of the density field and expansion coefficients can be very easily obtained,

$$\rho_{lm}(k) = \frac{i^l}{4\pi} \int \delta_{\mathbf{k}} Y_{lm}^*(\hat{\mathbf{k}}) d^2 \Omega_k \qquad (3-13)$$

where the well known relation,

$$e^{i\mathbf{k}\cdot\mathbf{r}} = 4\pi \sum_{lm} i^l j_l(kr) Y_{lm}(\hat{\mathbf{r}}) Y_{lm}^*(\hat{\mathbf{k}})$$

has been used.

The last two equations yield the auto-correlation function of the coefficients *

$$\langle \rho_{l_1 m_1}(k_1) \, \rho_{l_2 m_2}^*(k_2) \rangle = \frac{\pi}{2} \frac{P(k_1)}{k_1^2} \delta_{l_1,l_2}^K \delta_{m_1,m_2}^K \delta_D(k_1 - k_2) \qquad (3-14)$$

where P(k) is the usual power spectrum. This equation shows that this basis is statistically orthogonal and therefore, might be very useful for WF.

Now consider a 3-D flux limited catalogue, which suffers also from the incomplete sky coverage. Namely, the observed density is related to the real density $\rho$ by, $M(\hat{\mathbf{r}}) \, \phi(r) \, \rho(\mathbf{r})$, where $M$ is the mask function and $\phi$ is the radial selection function (we restrict the treatment to a pure radial selection function). The 'observed' expansion coefficients of the overdensity are:

$$\rho_{lmn}^{obs} = \int M(\hat{\mathbf{r}}) \, \phi(r) \, \omega(r) \, \delta(\mathbf{r}) \, j_l(kr) \, Y_{lm}(\hat{\mathbf{r}}) \, d^3 r \qquad (3-15)$$

where $\omega(r)$ is an arbitrary weighting function (often taken to be $\phi(r)^{-1}$).

---

* This equation obtained by using the orthogonality of spherical Bessel functions:

$$\int_0^\infty r^2 j_l(k_1 r) j_l(k_2 r) dr = \frac{\pi \delta_D(k_1 - k_2)}{2k_1^2}$$

The overdensity MV estimator $\hat{\rho}_{lmn}$ given a model is (equation 2-6):

$$\hat{\rho}_{lmn} = \sum_{l_1 m_1 k_1} \sum_{l_2 m_2 k_2} \langle \rho_{lmn} \rho^{obs,*}_{l_1 m_1 k_1} \rangle \langle \rho^{obs}_{l_1 m_1 k_1} \rho^{obs,*}_{l_2 m_2 k_2} \rangle^{-1} \rho^{obs}_{l_2 m_2 k_2} \qquad (3-16)$$

here the correlation function of the 'observed' coefficients is:

$$\langle \rho^{obs}_{l_1 m_1 1} \rho^{obs,*}_{l_2 m_2 2} \rangle = \int d^3 r_1 d^3 r_2 \, \phi(r_1)\phi(r_2)\omega(r_1)\omega(r_2) j_{l_1}(k_1 r_1) j_{l_2}(k_2 r_2)$$
$$\times Y^*_{l_1 m_1}(\hat{\mathbf{r}}_1) Y_{l_2 m_2}(\hat{\mathbf{r}}_2) M(\hat{\mathbf{r}}_1) M(\hat{\mathbf{r}}_2) \xi(|\mathbf{r}_1 - \mathbf{r}_2|)$$
$$+ \int d^3 r \, \phi(r)\omega^2(r) j_{l_1}(k_1 r) j_{l_2}(k_2 r) Y^*_{l_1 m_1}(\hat{\mathbf{r}}) Y_{l_2 m_2}(\hat{\mathbf{r}}) M^2(\hat{\mathbf{r}}) \qquad (3-17)$$

where $\xi$ is the correlation function. The last expression contains two terms, the first comes from the correlations and the second comes from the shot noise. Note that in the shot noise term $\phi$ appears to first power, while in the mask term it appears squared. This due to the different effect of the radial selection function on the noise. The first term at the RHS of equation (3-17) can be written, after some algebra, as:

$$I = \frac{2}{\pi} \sum_{l' m'} W^{m_1 m'}_{l_1 l'} W^{m_2 m'}_{l_2 l'} \int dk \, k^2 P(k)$$
$$\times \left[ \int dr_1 r_1^2 \phi(r_1)\omega(r_1) j_{l_1}(k_1 r_1) j_{l'}(k r_1) \right]$$
$$\times \left[ \int dr_2 r_2^2 \phi(r_2)\omega(r_2) j_{l_2}(k_2 r_2) j_{l'}(k r_2) \right] \qquad (3-18)$$

and the second term is:

$$II = \left(\frac{2}{\pi}\right)^2 \sum_{l' m'} W^{m_1 m'}_{l_1 l'} W^{m_2 m'}_{l_2 l'} \int dr \, r^2 \phi(r)\omega^2(r) j_{l_1}(k_1 r) j_{l_2}(k_2 r) \qquad (3-19)$$

The correlation function between the 'theoretical' and 'observed' coefficients is:

$$\langle \rho_{lmn} \rho^{obs,*}_{l_1 m_1 1} \rangle = \int d^3 r d^3 r_1 \phi(r_1)\omega(r_1) j_{l_1}(k_1 r_1) j_l(kr) M(\hat{\mathbf{r}}_1) Y^*_{lm}(\hat{\mathbf{r}}) Y_{l_1 m_1}(\hat{\mathbf{r}}_1) \xi(|\mathbf{r} - \mathbf{r}_1|)$$
$$(3-20)$$

This expression does not include a shot noise term. After some algebra, equation 3-20 can be written

$$\langle \rho_{lmn} \rho^{obs,*}_{l_1 m_1 k_1} \rangle = \frac{2}{\pi} W^{m_1 m}_{l_1 l} P(k) \int dr_1 r_1^2 j_{l_1}(k_{k_1} r_1) j_l(k r_1) \phi(r_1)\omega(r_1) \qquad (3-21)$$

some of the expressions become very simple when $\omega(r) = \phi(r)^{-1}$. Note, that even in this case equation (3-17) is not diagonal.

### 3d. Cartesian Coordinates

The last section shows that working in a 'statistically orthogonal' representation does not necessarily simplify the calculations, moreover, it turns out that if the ZOA covers more than $|b| = 15°$ (as is the case for most optical catalogues), then spherical harmonics reconstruction requires additional regularization beyond the WF (LFHSZ and section 5 below). In fact in some cases it might be more useful to perform the calculation in the real space (Hoffman 1994a,b). Here, MV estimator of the density is calculated in terms of Cartesian coordinates, where the considered model is an *IRAS* mock catalogue, extracted from a numerical simulation of a CDM Universe. The simulations are those used by Bistolas, Zaroubi and Hoffman (1994) and are of standard biased CDM model ($h = 0.5$, $\Omega = 1$, $\Lambda = 0$, $n = 1$). The output time of the simulation corresponds to an *rms* amplitude of mass fluctuations in an $8\,Mpc/h$ sphere of $\sigma_8 = 1$. The points selected as galaxies, however, are not biased; they do trace the mass, *i.e.*, $b = 1$ and $\Omega^{0.6}/b$ is unity. The mock catalogue was constructed to mimic the *IRAS* 1.2 Jy selection function. Here the ZOA modeled as a sharp mask at Galactic latitude $|b| = 5°$. For this case the following matrices are substituted in equation 2-6:

$$<s_i\, d_j> = \xi(|\mathbf{r}_i - \mathbf{r}_j|) \qquad (3-22a)$$

$$<d_j\, d_k> = \xi(|\mathbf{r}_j - \mathbf{r}_k|) + \frac{1}{\bar{n}(2\pi\sigma^2)^{3/2}} \int \phi^{-1} \exp\left\{-\left[\frac{(\mathbf{r}_i - \mathbf{x})^2 + (\mathbf{r}_k - \mathbf{x})^2}{2R_s^2}\right]\right\} d^3x$$
$$(3-22b)$$

where $\phi$ is the *IRAS* selection function (Yahil *et al.* 1991), $R_s = 10 Mpc/h$ and $\bar{n}$ is the mean number density of the mock catalogue. It is straight forward to calculate the shot noise term in equation 3-22b (Scherrer & Bertschinger 1991; Bistolas, Zaroubi and Hoffman 1994).

Figure 4[†] shows a reconstruction of the density field in the supergalactic plane from the mock *IRAS* CDM catalogue. The reconstruction is performed within a sphere of radius $r = 60 Mpc/h$ where 834 constraints are taken from radius of $40 Mpc/h$. Figure 4a shows the underlying field from which the catalogue is constructed. The 'observed' map smoothed over $10 Mpc/h$ is shown in Figure 4b. The MV reconstructed map is shown in Figure 4c. The contours are spaces at $\Delta\delta = 0.1$ with solid (dashed) line denoting positive (negative) contours. The heavy line represents $\delta = 0$, all the maps have been smoothed over $10 Mpc/h$. Figure 4d shows the reconstructed density within $r < 40 Mpc/h$ versus the true density as given by the N-body simulation.

## 4. Dynamical Reconstruction

The applications of the WF approach presented so far have all dealt with the statistical reconstruction of an underlying field from observational data that sample that field. In particular we have focused on the density field which is sampled by the galaxy distribution. Here, the reconstruction method is extended further to do dynamical reconstruction, namely using observational data that sample one field to reconstruct a different field that is dynamically related. Immediate application of that approach to the study of LSS includes the reconstruction of underlying density field from observed radial peculiar velocities, or going the other way round to use density data to construct the peculiar velocity field. This dynamical approach depends on the availability of a theoretical model which relates the two different fields *via* a cross-correlation function. Most of the discussion here is based on the linear theory of gravitational instability, however it will be shown how the formalism can be extended beyond the linear regime.

The velocity-density relation plays a crucial role in the study of LSS (for a recent review see Dekel, 1994). An example of an algorithm for reconstructing the density from the velocity field is the POTENT method where the observed radial velocities are integrated along the line-of-sight to get the velocity potential (Bertschinger and Dekel 1989; Dekel, Bertschinger and Faber 1990). This method assumes that the velocities are derived from a potential flow; with the additional assumption of linear theory the Laplacian of the potential yields the density fluctuations up to an overall proportionality constant. The drawback of this method, as is true with many other methods, is that it does not separate the signal (in this case the density field) from the noise (velocity errors). Hence velocity measurement errors are fed into the density determination. An indirect density-from-velocity reconstruction was done before by Kaiser and Stebbins (1991) and Stebbins (1994), who have defined the problem within the framework of Bayesian statistics. Their approach is equivalent, of course, to the WF method, and it is repeated here for the sake of completeness. The optimal dynamical reconstruction, $\delta^{\mathrm{opt}}(\mathbf{r})$, is derived by cross-correlating the radial velocity, $u(\mathbf{r}) = \mathbf{v}(\mathbf{r}) \cdot \hat{r}$ with $\delta(\mathbf{r})$:

$$\delta^{\mathrm{opt}}(\mathbf{r}) = \Big\langle \delta(\mathbf{r})u_i \Big\rangle \Big\langle u_i u_j + \sigma_i^2 \Big\rangle^{-1} U_j \quad . \qquad (4-1)$$

---

[†] V. Bistolas has kindly provided us with this figure

Here the $U_i = u_i + \epsilon_i$ is the observed radial velocity and for simplicity we assume a diagonal error covariance matrix. Given an assumed power spectrum the auto- and cross-correlation functions are readily calculated. Note that within the linear theory the velocity and density field are related by a simple convolution, and therefore the problem of density reconstruction from observed velocity is equivalent to a deconvolution in the presence of noise (with possible predictions to unobserved regions). The inverse problem of estimating the velocity field from the measured galaxy distribution is very similar to the above velocity reconstruction.

The density-density reconstruction has been given in §3. The velocity-density relation is analogous to equation 4-1, i.e.,

$$\mathbf{v}^{\mathrm{opt}}(\mathbf{r}) = \Big\langle \mathbf{v}(\mathbf{r})\delta(\mathbf{r}_i) \Big\rangle \Big\langle \delta(\mathbf{r}_i)\delta(\mathbf{r}_j) \Big\rangle^{-1} \Delta(\mathbf{r}_j) \quad . \qquad (4-2)$$

A major problem in analyzing red-shift surveys is the transformation of the galaxies from redshift to real space. Two direct approaches to the problem have been used in analyzing the *IRAS* redshift catalog. One is the iterative method of Yahil *et al.* (1991), and the other more recent modified Poisson equation of Nusser and Davis (1994). The latter method relates the velocity potential to the density evaluated in redshift space, resulting in a Poisson equation with an extra term added to it. These two direct methods do not remove the noise before applying the dynamical mapping, but rely on smoothing to mitigate its effects. Within the limitation of the linear theory a WF can be used to remove noise, transform from redshift to real space, deconvolve smoothing and extrapolate across unobserved regions (*cf.*, FLHLZ). This is given by

$$\delta^{\rm opt}(\mathbf{r}) = \Big\langle \delta(\mathbf{r})\delta_S(\mathbf{s}_i) \Big\rangle \Big\langle \delta_S(\mathbf{s}_i)\delta_S(\mathbf{s}_j) \Big\rangle^{-1} \Delta_S(\mathbf{s}_j), \qquad (4-3)$$

and here **s** is the position vector in redshift space, subscript $S$ denotes quantities evaluated in redshift space and $\Delta_S$ is the observationally determined density (contrast) as evaluated in redshift space. The auto- and cross- correlation matrices which relate the redshift space and real space densities are readily calculated in the linear theory (*cf.*, Zaroubi and Hoffman 1994). In the linear theory the redshift space density depends linearly on the actual (real space) densities, and therefore one can write a linear transformation to relate the two and the WF approach amounts a matrix inversion in the presence of noise. Indeed, the redshift to real space transformation in the SH presentation is given in terms of a regularized inversion of a distortion matrix (FLHLZ).

### 5. Extensions

#### 5a. Extra-Regularization

In several cases the WF alone is not sufficient to stabilize the deconvolution and consequently it must be either modified, supplemented with extra-regularization, or replaced with another method. In this section we examine several examples of extra-regularization and briefly discuss other possible regularization techniques.

### 5a.1 SVD as a Regularizer

A simple example which illustrates the need for further regularization is the reconstruction of the angular galaxy distribution when the ZOA is quite large, as in the case of optical galaxy surveys. In this case the stability is controlled by four factors, namely the *prior* , the shot noise, the width of the mask and the required resolution (*i.e.*, $l_{max}$). Mathematically, the reconstruction of the full sky harmonics given in equation 3-9, involves the solution of the equation

$$\mathbf{B}\mathbf{a}^{MV} = \mathbf{C} \qquad (5-1)$$

for the unknown harmonic coefficients, $\mathbf{a}^{MV}$.

An elegant and robust solution to such a linear set of equations is given by the Singular Value Decomposition algorithm (SVD) (e.g., Press *et al.* 1992) The SVD algorithm basically decomposes any $M \times N$ matrix, $\mathbf{B}$, into a multiplication of three matrices, $\mathbf{B} = \mathbf{U}\ diag\{\lambda_i\}\ \mathbf{V}^T$ where the set $\{\lambda_i\}$ are the referred to as the singular values of the matrix $\mathbf{B}$. The matrices $\mathbf{U}$ and $\mathbf{V}$ are each orthogonal in the sense that their columns are orthonormal. The matrix $\mathbf{U}$ is also of order $N \times M$ while the other two are square $N \times N$ matrices. The inversion, after the decomposition, is straightforward and gives,

$$\mathbf{B}^{-1} = \mathbf{V}\ diag\{1/\lambda_i\}\ \mathbf{U}^T. \qquad (5-2)$$

Formally speaking, equation 5-1 has a unique solution if and only if $\mathbf{B}$ is a non-singular matrix, namely if $\lambda_i \neq 0$ for all $i$. However a meaningful solution to equation 5-1 can be obtained even in the case where $\mathbf{B}$ is singular, by requiring the solution to minimize the norm of the residuals, $|\mathbf{B}\mathbf{a} - \mathbf{C}|$. Such a solution is obtained by substituting $1/\lambda_i = 0$ in the expression for the inverse (equation 5-2) for any $\lambda_i = 0$ (Press *et al.* 1992).

The question arises in a particular problem of setting the lower limit of the singular values, below which the inverse values are set to zero. As an example, we consider the problem of reconstructing the harmonic coefficients for the simulations discussed in § 3b when the ZOA is quite large, $|b| \leq 15°$, (characteristic of optical galaxy catalogues). Figure 5 shows the harmonic reconstruction of the projected counts of the simulation using the raw harmonic coefficients, $c_{lm}$, up to $l_{max} = 15$. In this case the direct inversion of the matrix **B** is unstable and yields excessive power on small scales. However using the SVD algorithm sheds a new light on the question of the stability of the WF reconstruction. Figure 6 shows the sorted spectrum of the singular values ($\lambda_i$) versus the harmonic number, $l$. In general, the singular values measure the amount of 'information' carried by each mode in the problem (Press *et al.* 1992), namely the small singular values does not have significant contribution to reconstruction, nevertheless, they can destabilize the inversion. As an extension of the ideal case of $\lambda_i = 0$, we impose a cutoff on the small singular values in order to maintain stability. The 'knee' in Figure 6 suggests how to choose the cutoff for the SVD, where in this case it is $\lambda_{min}/\lambda_{max} = 0.565$. Note that this cutoff suggests that only the largest 184 modes are significant, corresponding to an effective $l_{max} \approx 13$. This implies that with a mask of $|b| = 15°$ (and assumed shot-noise and power spectrum) there in no need to expand the harmonics much beyond $l = 13$.

Figure 7 shows the the reconstructed $a_{lm}$'s map using WF and the above cutoff for SVD. Note that here, contrary to the $|b| = 5°$ case discussed by LFHSZ where the structure was recovered in the ZOA, the reconstructed ZOA remains empty, which illustrates that WF even with the use of SVD extra-regularization can not create structure out of nothing, unless it is dictated by the correlations (*cf.*, LFHSZ).

This example shows how by using the singular values one can gain more insight to physical relevance of the harmonic modes and how to use that to stabilize the deconvolution operation. Yet, as with the WF, the application of the SVD does not guarantee the stability of reconstruction. In cases where the amount of observable data, its quality and the nature of the *prior* do not set strong enough constraints on the underlying field other regularization techniques might be used.

**5a.2 Maximum Entropy**

The MaxEnt algorithm is closely related to the Bayesian approach discussed in § 2c and it can actually be regarded as a particular application in which the PDF of the *prior* is given as the exponent of some regularization functional, usually called 'entropy' functional (Narayan and Nityananda 1986). The common application of the MaxEnt technique is in image reconstruction, where one often deals with very high signal-to-noise images. In such cases the entropy is formulated in general terms of information theory without **any** reference to the physical nature of the object whose image is studied. Thus an image of an Sb galaxy taken by an astronomical telescope or a picture of New York City taken from an Earth-orbiting satellite are analyzed by the same MaxEnt technique (*e.g.*, Puetter and Pina, 1993). The analysis of LSS presents a very different challenge, involving data sets whose typical signal-to-noise ratio is rather low, of the order of a few or even unity. However, LSS has been extensively studied and a lot of information has been accumulated. This information should be taken into account in constructing a *prior* model, and this should compensate for the low quality of the data. Thus, unlike the case of image reconstruction where the entropy describes the information content of the image regardless of the physical nature of the object at hand, a MaxEnt approach to LSS should be based on an entropy formulation which depends on the physics of the underlying field. Applications of the MaxEnt method for image restoration in astronomy, using various entropy functionals has been reviewed by Narayan and Nityananda (1986).

Maximum Entropy can be viewed as a regularization procedure where the quantity maximized is

$$Q = -L + \alpha\, S$$

where $(-L)$ is the log-likelihood function $(-\frac{1}{2}\chi^2)$, $\alpha$ is a Lagrange multiplier, and $S$ is the entropy. When considering reconstructions of cosmological fields it seems reasonable to choose the galaxy density $\rho$ as the relevant variable, and to adopt Skilling's generalization of Shannon's entropy (Skilling 1988) :

$$S(\rho) = \rho - \bar{\rho} - \rho \ln(\rho/\bar{\rho}),$$

such that when $\rho = \bar{\rho}$ the entropy is zero, and the form ensures that the density cannot be negative. Maximizing the entropy under the data constraint (likelihood) will give us the most conservative picture of deviations from uniformity allowed by the data. For a similar application of reconstructing the galaxy density field with Maximum Entropy see Lahav and Gull (1989).

We now note that for small fluctuations $\delta = (\rho - \bar{\rho})/\bar{\rho}$, the entropy is $S \approx -\frac{1}{2}\bar{\rho}\delta^2$. This quadratic form as a regularizer (which does not ensure positivity) leads a Wiener filter (as described in § 2). Hence in the case of small fluctuations the Maximum Entropy and Wiener approaches are very similar. We note that in our Wiener approach the Lagrange multiplier is not a free parameter; it is fixed in terms of the assumed cosmic variance.

### 5b. Alternative Filters

The WF is based on a minimal variance approach, and therefore it should provide a powerful reconstruction technique for random systems whose statistical behavior does not depend on moments higher than the second, namely the variance. Not surprisingly it has been shown here that in the case of GRFs the WF coincides with MAP, MV and mean estimators of the underlying field. Thus, to the extent that the system under consideration is well described by the formalism of GRFs one might reasonably expect the WF to perform well. The case of the linear regime of LSS certainly falls in this category.

As mentioned before, the WF has its limitations, even in the framework of GRFs. It depends on an adequate modeling of the data, and an assumed knowledge of the correlation function of the underlying field and the nature of the statistical uncertainties. Now, in the limit of no statistical uncertainties, the WF reduces to a simple deconvolution and prediction. This can be extended over an infinite domain (in real space) or to infinite resolution (e.g., in Fourier space). The existence of noise limits the extrapolation of the reconstruction in the two conjugated spaces. Yet, the tendency of the WF is to vanish in the absence of good data, yielding the null field as the best estimator. The WF always yields a conservative estimate of the underlying field, replacing noise by the zero field. An *ad hoc* attempt to correct for this conservative approach is given by a modified WF (*cf.*, Andrews and Hunt, 1977), which in the case of a diagonal WF has a simple form. Here we apply it to the case of Fourier space reconstruction involving smoothing, where the point spread function is $W(k) = \exp(-k^2 R^2/2)$ ($R$ is the smoothing length). A modified filter yields,

$$\delta_{\mathbf{k}}^{\mathrm{est}} = \Big(\frac{P(k)W(k)}{P(k)W^2(k) + \sigma^2}\Big)^{1-\Gamma} W^{-\Gamma}\Delta_{\mathbf{k}}, \qquad (5-3)$$

where $\Gamma$ ranges from 0 to 1. Note that in the limit of vanishing noise ($\sigma^2$) one obtains simple deconvolution, which is also the case of $\Gamma = 1$. The standard WF is recovered for $\Gamma = 0$. Adopting values of $\Gamma$ larger than 0, increases the amplitude of the filter and thus the amplitude of the estimated field. The reader should be aware that this stronger signal is obtained at the expense of weaker statistical significance (larger variance) since more of the observational noise is retained as "real" signal. A recent application of a modified WF has been done by A. Yahil (private communication)

### 5c. Non-Gaussian Fields

As the perturbation field evolves to the non-linear regime it ceases to be a GRF, namely its (Fourier) phases become correlated. Much of the success of the application of the WF to the reconstruction of the LSS is due to the Gaussian nature of the primordial perturbations. The question naturally arises as to what extent the WF approach is applicable beyond the linear regime. Here we present two different approaches, dynamical and statistical, to the problem.

It has been argued that the statistical properties of the perturbation field in the quasi-linear regime are well approximated by a log-normal distribution (Coles and Jones 1991; Sheth 1994). Given that, the WF can be applied to the reconstruction of the logarithm of the density to yield an optimal estimator of the $log[\rho]$ field. This can be easily done either by estimating the correlation functions directly from the observations or by calculating them theoretically from a chosen model. However, ambiguity arises when this is translated to the actual density field. For a log-normal distribution the different estimators, namely the optimal (in least squares sense), mean and most probable, do not coincide (Sheth 1994). Also, special care should be given to the treatment of errors in the determination of the $log[\rho]$ field. Note however that to the extent that the log-normal distribution is a good fit to the data, the WF provides a statistical rigorous reconstruction tool and the ambiguity in its application arises from the nature of the assumed log-normal statistics. Note also that such a PDF preserve the positivity of the density field.

A very different approach is to perform the reconstruction on the density field itself, namely applying the WF as a minimal variance estimator regardless of the actual statistical distribution. Now, the nature of the gravitational instability is that the fractional overdensity ($\delta$) cannot be negative, and therefore as the dynamics evolves away from the linear regime skewness develops (Bouchet *et al.* 1993, Nusser, Dekel & Yahil 1994). The WF minimizes the variance and it tends to ignore the skewed nature of the distribution. Thus, at least in the form presented here, the WF is expected to yield poor reconstruction of the LSS in the very non-linear regime. Yet it might be of some use in the quasi-linear regime. This, rather naive, application of the WF can be improved considerably in the following way. Consider the problem of the reconstruction of the primordial perturbation field from the present day quasi-linear observed density and/or velocity field, *i.e.*, the recovery of the linear field from the present epoch quasi-linear structure. Formally this can be obtained by using equation 4-3 in which the index $S$ is substituted by $QL$, which refers the quasi-linear density (or velocity) field. This depends on the evaluation of the auto- and cross-correlation functions. This can be achieved by direct analytical calculation within the framework of perturbation theory, or evaluated empirically from fully non-linear numerical simulations. Note that in this case the minimal variance reconstruction coincides with the mean field and most probable Bayesian estimator (within the framework of the GRFs). To reconstruct the present day LSS one can use CR to add short waves power and set initial conditions for N-body simulation. Integrating the equations of motion one obtain a full non-linear reconstruction of the LSS.

Finally, in the limit of small density number of galaxies in the catalogue, the shot noise is no longer well represented by a Gaussian field. While, the Gaussian assumption is reasonable for most regions, a rigorous treatment of the WF in underdense regions requires the noise to be described by a proper Poisson PDF.

### 5d. Simultaneous Reconstruction and Parameters Estimation

Suppose that the physical model one uses to construct the *prior* has a few free parameters. Generally, one would like to determine the parameters in the *prior* as well as the underlying field using the same data. The Bayesian approach provides one with a method of simultaneously determining the field and the parameters. In the context of our problem, there are two kinds of parameters, internal and external. The internal parameters are free parameters appearing in the theoretical *prior* (e.g. $\Omega$ and the bias parameter). while the extrenal parameters account for possible corrections to the model (e.g. the effect of matter outside the sampled volume on the reconstructed peculiar velocities). Equation 2-1 can be readily extended to include a set of linear external parameters, $\mathbf{q}$,

$$\mathbf{d} = \mathbf{R}\mathbf{s} + \mathbf{T}\mathbf{q} + \boldsymbol{\epsilon}. \qquad (5-4)$$

where the linearity is represented by the $M \times L$ matrix $\mathbf{T}$. (The estimation of the external parameters was analysed by Rybicki and Press,1992, and this is generalized here to the case of internal parameters as well)

A Bayesian approach to the problem is to write the *posterior* PDF,

$$P[\mathbf{s},\mathbf{p},\mathbf{q}|\mathbf{d}] = \frac{P[\mathbf{s},\mathbf{p},\mathbf{q}]P[\mathbf{d}|\mathbf{p},\mathbf{q},\mathbf{s}]}{P[\mathbf{d}]} \propto P[\mathbf{s}]P[\mathbf{d}|\mathbf{s}] \quad , \qquad (5-5)$$

where we have included the dependence of the *prior* of a set of internal parameters, $\mathbf{p}$. The last proportionality in equation 5-5 holds for the case of an assumed uniform *prior* for **p** and **q** ($P[\mathbf{d}]$ serves only to normalize the probability function). Extending equation 2-15 to include the free parameters, one finds

$$P[\mathbf{s},\mathbf{p},\mathbf{q}|\mathbf{d}] \propto \sqrt{\frac{1}{\det(\mathbf{S})}} \times \quad (5-6)$$

$$\exp\left\{-\frac{1}{2}\left[\mathbf{s}^\dagger \mathbf{S}^{-1}\mathbf{s} + \left(\mathbf{d}-(\mathbf{Rs}+\mathbf{Tq})\right)^\dagger \mathbf{N}^{-1}\left(\mathbf{d}-(\mathbf{Rs}+\mathbf{Tq})\right)\right]\right\}$$

$$\equiv Z \exp\{-\frac{1}{2}\theta\}$$

Note that here the correlation matrix, **S** depends on **p** while **N** does not. The second term in the argument of the exponent ($\theta$) is the $\chi^2$ of the data for a given set of parameters and a realization of he underlying field:

$$\theta = \mathbf{s}^\dagger \mathbf{S}^{-1}\mathbf{s} + \chi^2 \quad (5-7)$$

The Bayesian, namely MAP, simultaneous estimation of the field and parameters is obtained by solving for the extrema of the *posterior* PDF in the multidimensional $(N+L+P)$ space:

$$\left[\frac{\partial}{\partial \mathbf{s}}, \frac{\partial}{\partial \mathbf{p}}, \frac{\partial}{\partial \mathbf{q}}\right] P[\mathbf{s},\mathbf{p},\mathbf{q}|\mathbf{d}] = 0 \quad (5-8)$$

Consider first the estimation of **q**. The only external dependence of the *posterior* PDF is in the $\chi^2$, and therefore the vanishing of the gradient in the subspace of the external parameters implies that

$$\frac{\partial}{\partial \mathbf{q}}\chi^2 = 0. \quad (5-9)$$

Solving this equation one finds (RP):

$$\mathbf{q}^{\mathrm{MAP}} = \left(\mathbf{T}^\dagger \mathbf{D}^{-1}\mathbf{T}\right)^{-1} \mathbf{T}^\dagger \mathbf{D}^{-1}\mathbf{d} \quad (5-10)$$

Next, one takes the first derivative of equation 5-6 with respect to **s**. The MAP solution is equivalent to the WF, and upon substituting $\mathbf{q}^{\mathrm{MAP}}$ one finds:

$$\mathbf{s}^{\mathrm{MAP}} = \mathbf{S}\mathbf{R}^\dagger \left[\mathbf{R}\mathbf{S}\mathbf{R}^\dagger + \mathbf{N}\right]^{-1} \left(\mathbf{d} - \mathbf{T}\mathbf{q}^{\mathrm{MAP}}\right) \quad (5-11)$$

Finally, the internal parameters are to be determined here, thus solving

$$\left[\frac{\partial P[\mathbf{s},\mathbf{p},\mathbf{q}|\mathbf{d}]}{\partial \mathbf{p}}\right]_{\mathbf{q}^{\text{MAP}},\mathbf{s}^{\text{MAP}}} = 0. \qquad (5-12)$$

Now, given equation 2-16 it can be easily shown that on the subspace where $\mathbf{q} = \mathbf{q}^{\text{MAP}}$, $[\theta] = 0$. (*here it might be equal to an irrelevant constant, independent of data.*) Next one finds,

$$\left[\frac{\partial \theta}{\partial \mathbf{p}}\right]_{\mathbf{q}^{\text{MAP}},\mathbf{s}^{\text{MAP}}} = \frac{\partial \mathbf{d}^{\dagger}\left(\mathbf{R}\mathbf{S}\mathbf{R}^{\dagger} + \mathbf{N}\right)^{-1}\mathbf{d}}{\partial \mathbf{p}} \equiv \frac{\partial \chi^2_{\text{CV}}}{\partial \mathbf{p}}, \qquad (5-13)$$

where $\chi^2_{\text{CV}}$ is the $\chi^2$ that takes into account the cosmic variance, *i.e.*, all possible realizations of the filed $\mathbf{s}$. The condition imposed by equation 5-9 now yields,

$$\frac{d}{d\mathbf{p}}\left(\log Z - \frac{1}{2}\chi^2_{\text{CV}}\right) = 0. \qquad (5-14)$$

Thus solving equation 5-12 one finds the most probable values of the parameters, $\mathbf{p}^{\text{MAP}}$, given the data. Note however that this is exactly the result obtained by writing the likelihood function of the data given the parametric model (after correcting for the external parameters),

$$\mathcal{L}(\mathbf{d}|\mathbf{p}) = Z \exp\left(-\frac{1}{2}\chi^2_{\text{CV}}\right). \qquad (5-15)$$

Namely a likelihood function which takes into account all possible variations of $\mathbf{s}$, *i.e.*, the cosmic variance. The conclusion that follows is that the likelihood analysis of the parameters is independent of the Wiener estimation, but is consistent with it.

## 6. Discussion

The general framework of linear estimation and prediction by minimal variance, also known as Wiener filtering, has proved to be a very useful tool for reconstructing the large scale structure of the universe from incomplete, noisy and sparse data. In particular this holds on scales where the linear theory is valid and the underlying perturbation field is Gaussian. In such a case the WF solution of minimal variance coincides with other estimators such as the Bayesian MAP solution, conditional probability and maximum entropy (for a quadratic entropy). In conjunction with the algorithm of constrained realization of GRFs, the combined WF/CR approach provides one with a method of reconstructing, predicting and performing Monte Carlo simulations of the LSS. Various applications of the method include the reconstruction of the angular (LFHSZ) and three dimensional (Hoffman 1994a,b) galaxy distribution, the peculiar velocity and density fields (FLHLZ), and the CMB large scale anisotropies (BFHLSZ). This approach is currently being used to set initial conditions for high resolution N-body simulations from low resolution *IRAS* data and to perform Bayesian parameter estimation from COBE DMR's data.

The WF has lost its appeal in many fields of physics and technology where one faces similar problem of reconstruction from noisy and blurred data, and in particular so in the field of image reconstruction. The field of cosmology and LSS, on the other hand, provides one with an ideal case for using the WF. The main shortcoming of the WF is that it involves only the first and second moments (mean and variance) of the statistical distribution, ignoring higher moments. However, a fundamental ingredient of the canonical model of cosmology and LSS is that the primordial perturbation field is Gaussian, *i.e.*, it is determined by its first two moments. Thus a WF would reconstruct all the key ingredients of such a field, and would not add any new ones as an artifact. Furthermore, the basic underlying symmetries of the cosmological model are homogeneity and isotropy, and this statistical homogeneity is a necessary condition for a successful application of the method. Thus, as long as one deals with structures in the linear regime, where the GRF approximation holds, the WF is the tool of choice for reconstruction. Yet one problem still remains, and that is related to the basic property of the filter to predict the null field in the absence of good data. In the limit of very poor signal-to-noise data and/or far away from regions of good data, the cosmological mean field is estimated, *i.e.*, zero perturbation. Now, this is what one expects of course, but this is not very useful for Monte Carlo and modeling purposes. A common astrophysical problem is that the quality of the data degrades with distance, and as the signal-to-noise ratio decreases the amplitude of the WF estimated field decreases with it. Yet the model tells us that the variance of the field is constant in space. This problem is solved here by making CR on the data, *i.e.*, adding to the minimal variance solution a random realization of the residual from the mean. The CR is guaranteed to have the correct variance, and where the quality and sampling of the data are good, the resulting CR is dominated by the data and the actual underlying field is recovered. Otherwise, almost random realizations are obtained that are dominated by the *prior* model, but still somewhat constrained by the available data.

The traditional WF approach focuses on recovering the underlying field, whose statistical behavior is determined by the *prior* model. However, commonly the cosmologist analyzing a data set is primarily interested in estimating the free parameters that determine the model and only then to estimate the field itself. The Bayesian approach, which in the case of GRFs coincides with the minimal variance solutions, provides one with a unified way of estimating both parameters of a model and a particular realization of the field. Having this, the generalized WF approach (*i.e.*, including CRs and *posterior* Bayesian likelihood analysis) seems to establish a comprehensive framework of analyzing the (linear) LSS structure of the universe. This should be further studied and extended to the non-linear regime of non-Gaussian random fields.

## Acknowledgments

We would like to especially thank V. Bistolas and C. Scharf for helping with the applications of the method. We thank E. Bunn, A. Dekel, D. Lynden-Bell and R. Sheth for helpful comments and discussions. KBF acknowledges a SERC postdoctoral fellowship. YH has been partially supported by The Hebrew University Internal Funds. YH and SZ acknowledge the hospitality of the Institute of Astronomy in Cambridge. OL acknowledges the hospitality of the Hebrew University, where this work began.34

**Figure Captions**

**Figure 1:** Harmonic reconstruction of the projected counts of an IRAS-like CDM simulation, using the raw all-sky coefficients $c_{lm}$ up to $l_{max} = 15$ in Galactic Aitoff projection. The contour levels of the projected surface number density are in steps of 100 galaxies per steradian. The mean projected density is 392 galaxies per steradian.

**Figure 2:** Wiener filtering of the all-sky harmonic map shown in Figure 1, assuming the standard CDM prior and the simulation shot noise (392 galaxies per steradian).

**Figure 3:** The Wiener factor versus the harmonic number $l$, where the Wiener factor is the ratio, $\frac{<a_l^2>_{th}}{<a_l^2>_{th}+<\sigma_a^2>}$ evaluated for a standard CDM model with the same selection function as $IRAS$ 1.2 Jy sample. The shot noise is $\sigma_a^2 = 392$ galaxies per steradian.

**Figure 4:** Density reconstruction of the Mock $IRAS$ CDM catalogue in the supergalactic plane. The reconstruction is performed within a sphere of radius $60 Mpc/h$. a) The underlying density field as taken from the N-body simulation. b) The 'observed' map. c) The MV reconstructed map. The contours are spaced with $\Delta\delta = 0.1$ with solid (dashed) lines denoting positive (negative) contours. The heavy solid line stands for $\delta = 0$. All the maps have been smoothed over $10 Mpc/h$. Figure d. is a scatter plot of the MV reconstructed density *versus* the true density within a sphere of radius $40 Mpc/h$.

**Figure 5:** Harmonics reconstruction of the projected counts of the Mock $IRAS$ 1.2 Jy catalogue with a 'Zone of Avoidance' $|b| = 15°$, using the observed coefficients $c_{lm}$, up to $l_{max} = 15$. The contour levels and the mean projected density are the same as in Figure 1.

**Figure 6:** The sorted spectrum of the singular values of the matrix **B**, for the $|b| = 15°$ case. The 'knee' suggests how to choose the cutoff for the SVD, $\lambda_{min}/\lambda_{max} = 0.565$, at the mode 184, which corresponds to an effective $l_{max} \approx 13$.

**Figure 7:** The reconstructed alm's using WF and the cutoff shown in Figure 6 for the SVD.

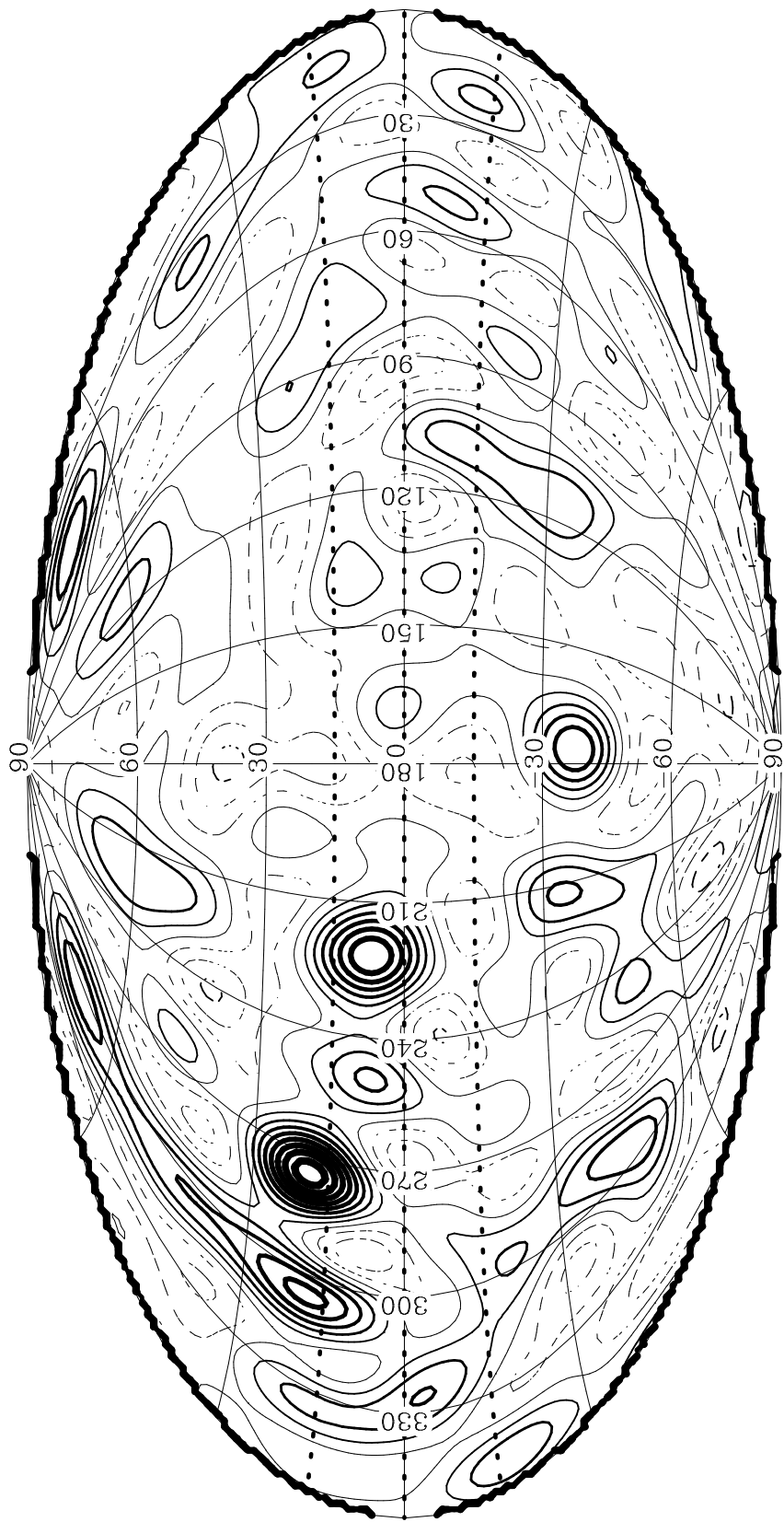

Figure 1

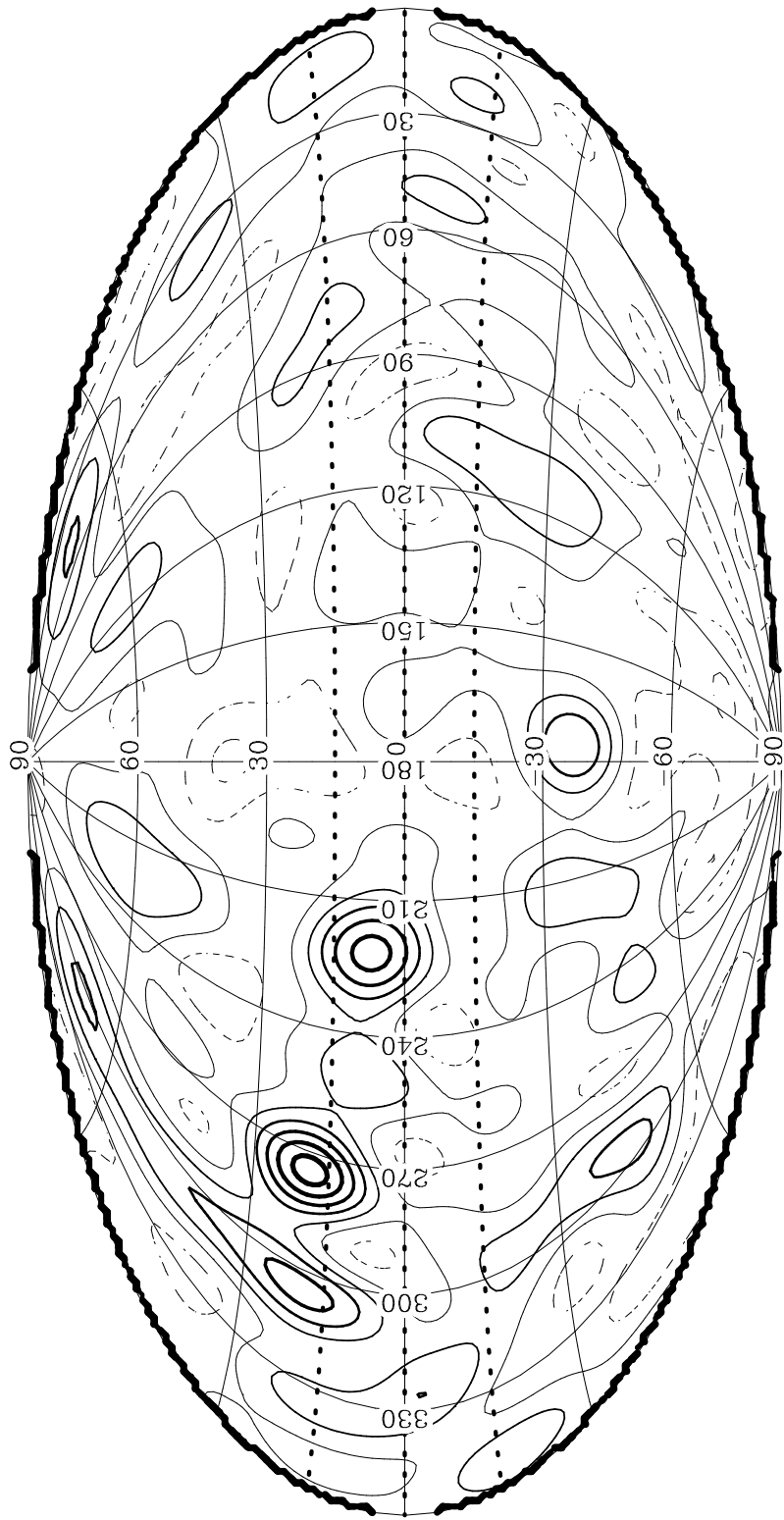

Figure 2

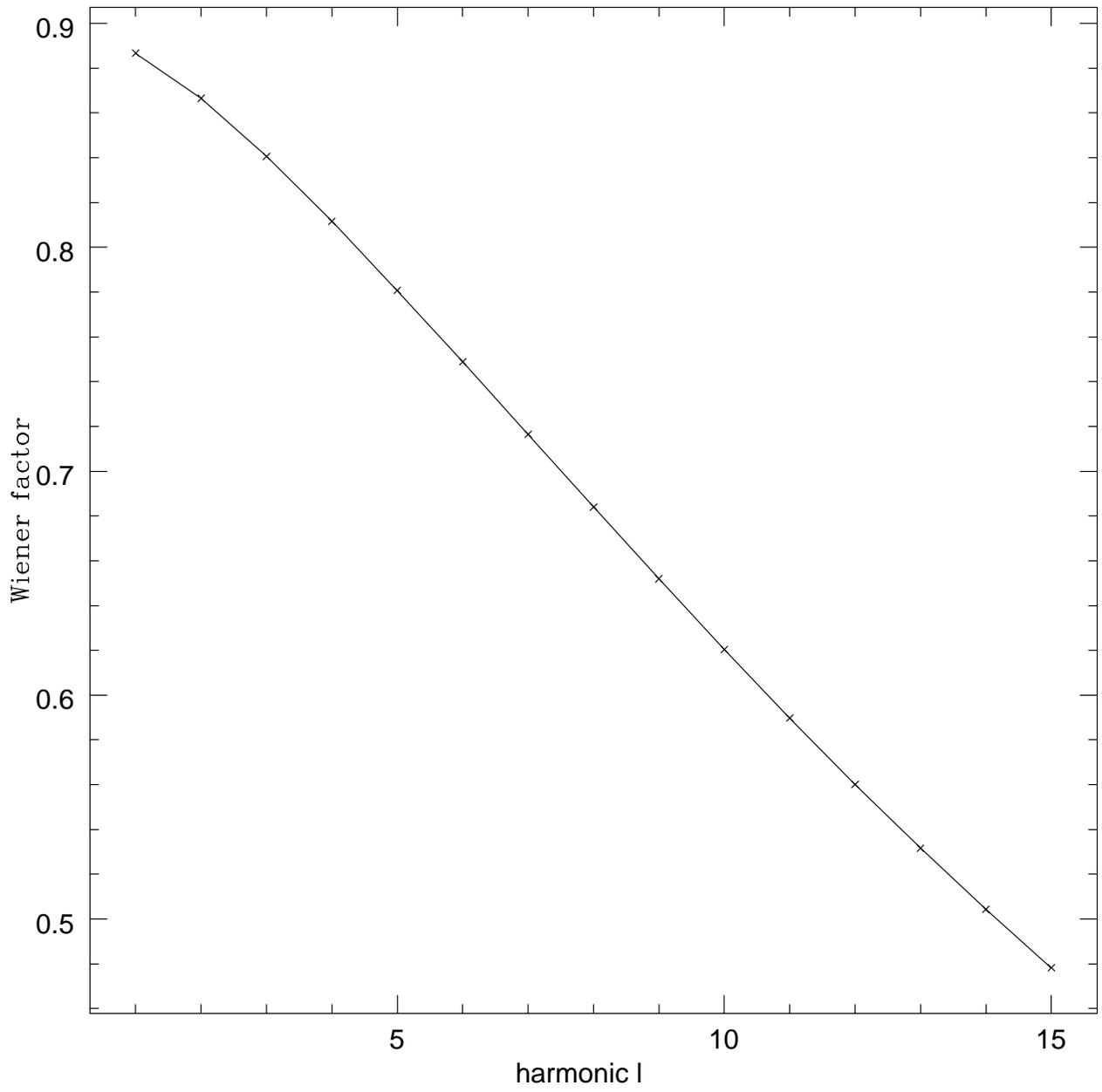

**Figure 3**

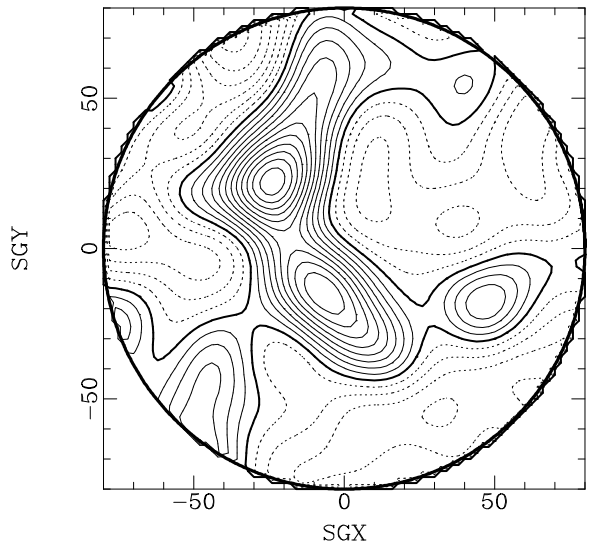
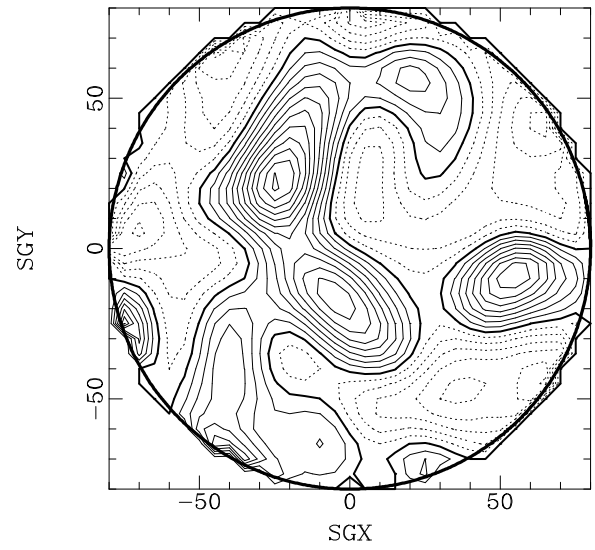
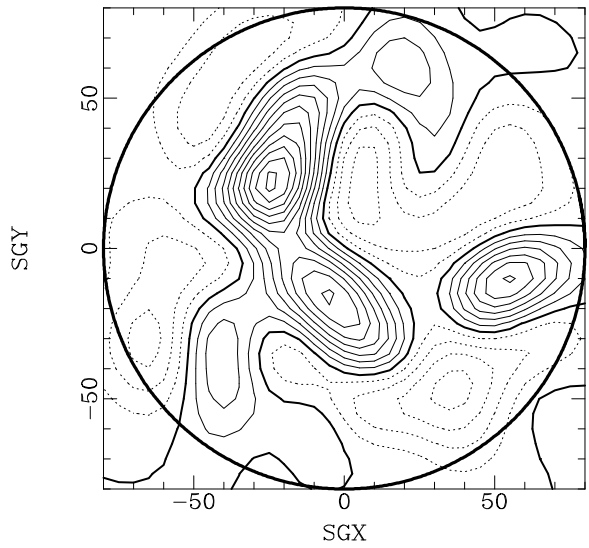
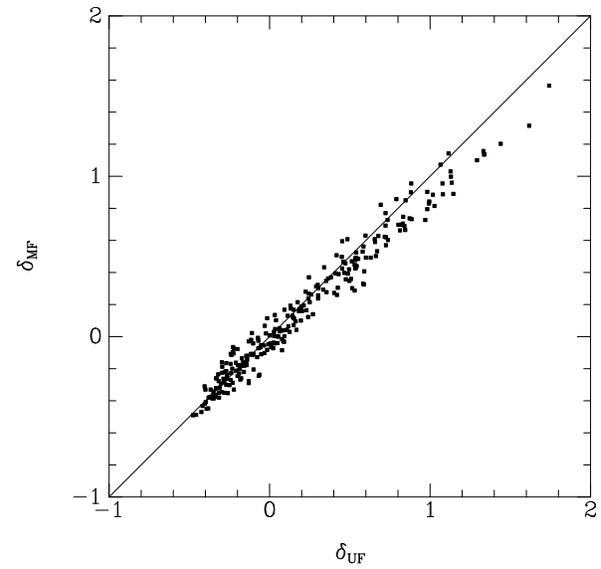

**Figure 4**

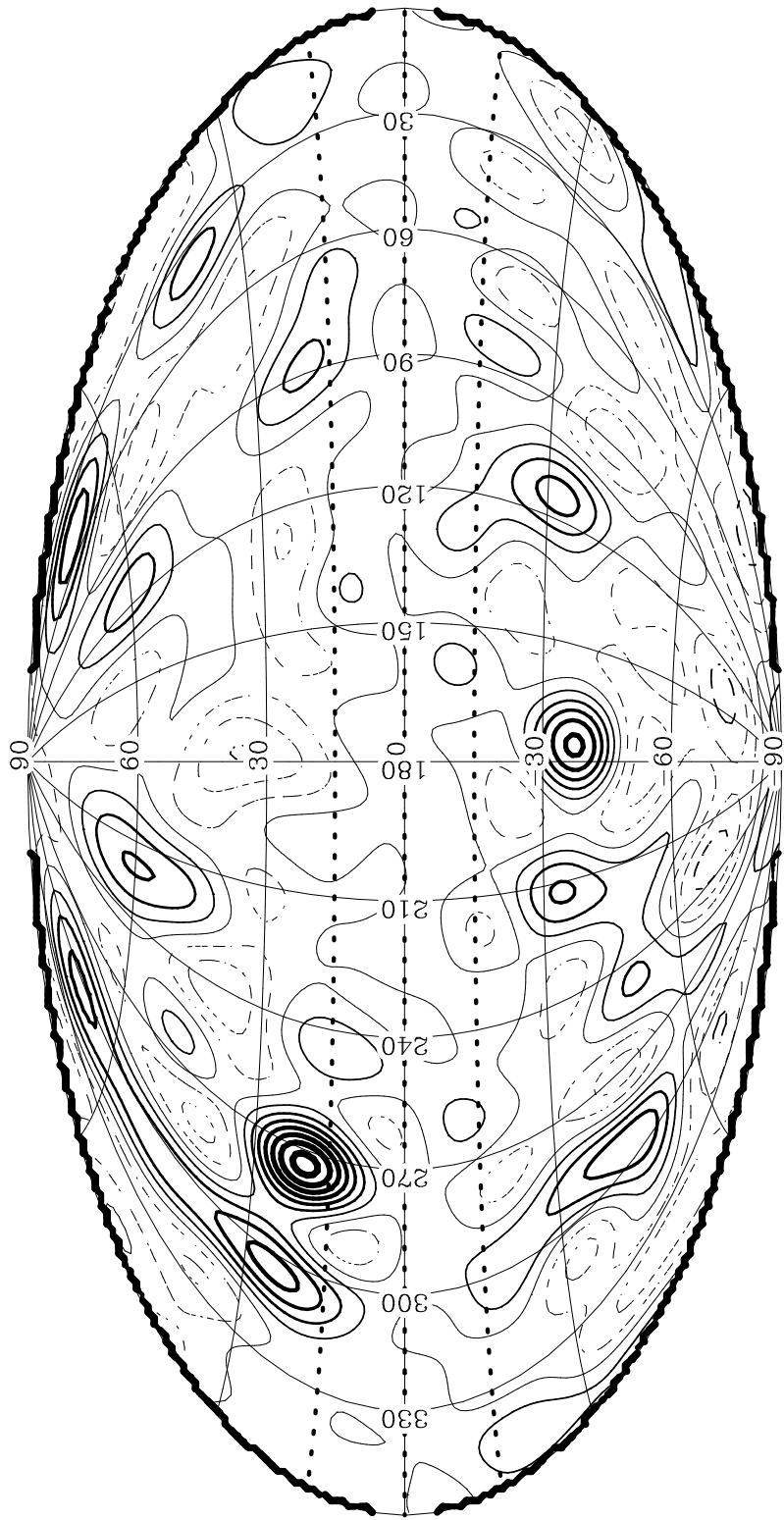

Figure 5

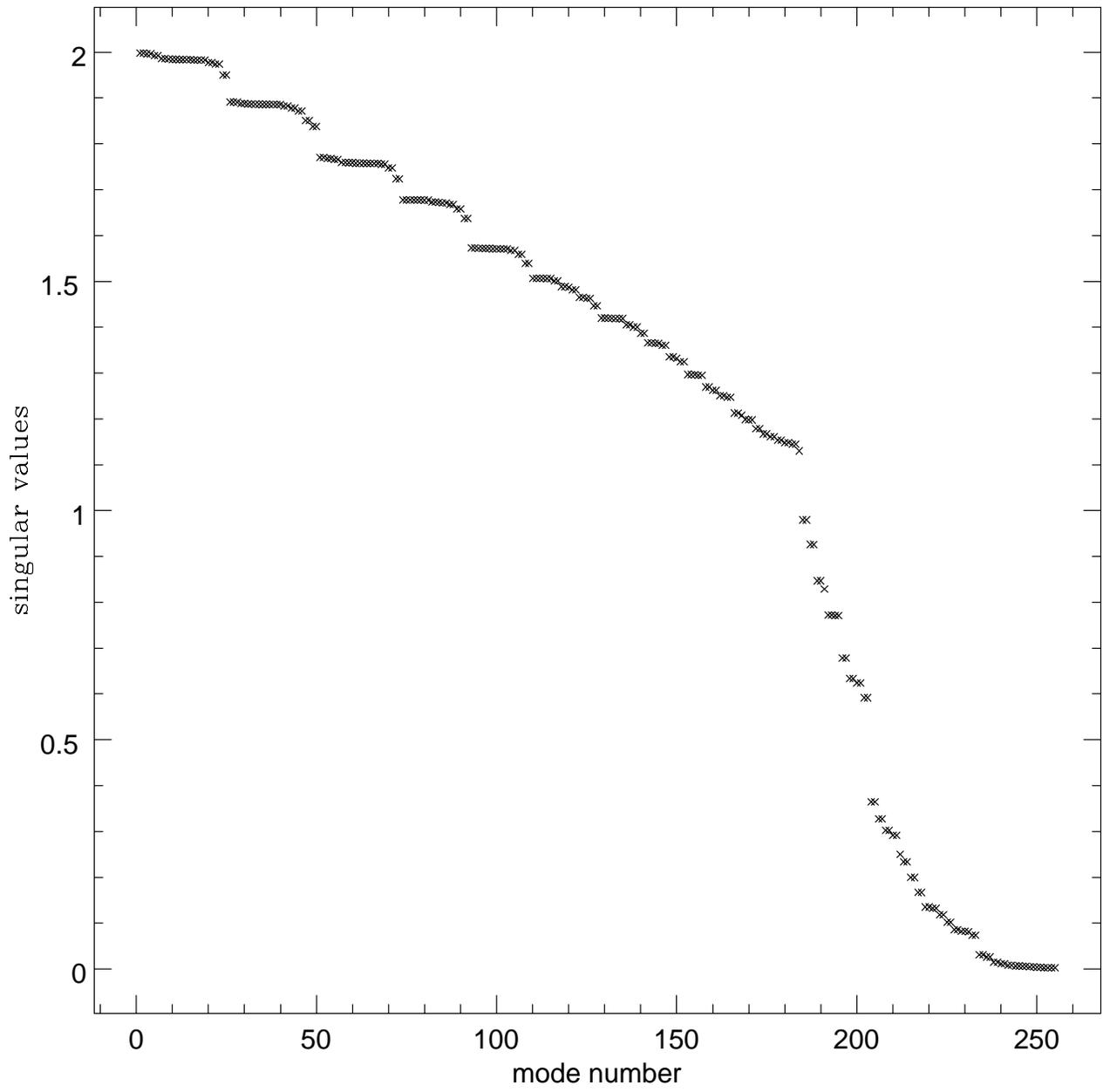

**Figure 6**

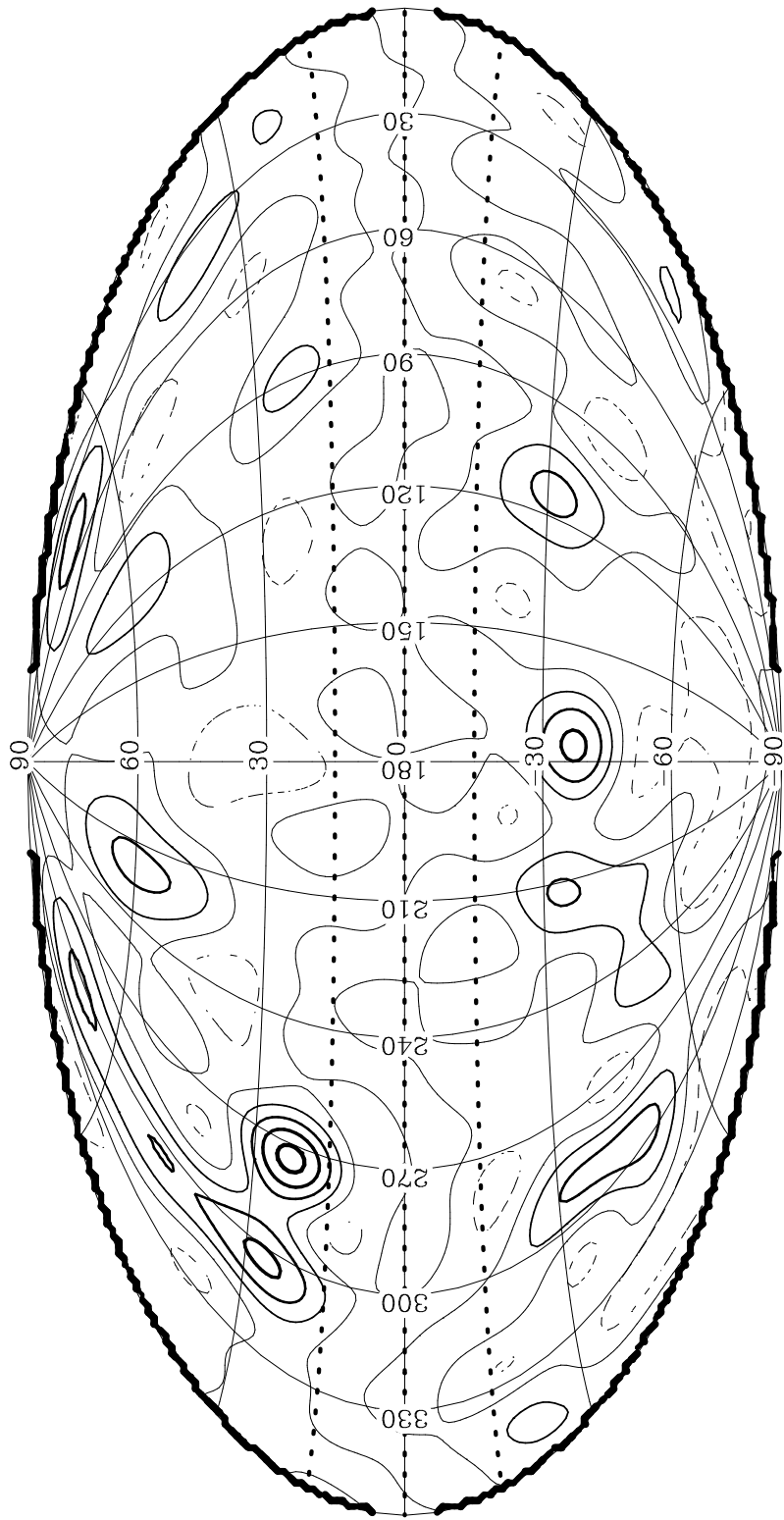

Figure 7